\documentclass[prd,12pt,nofootinbib,showpacs,tightenlines]{revtex4}

\usepackage[frenchb,english]{babel}
\usepackage[latin1]{inputenc}
\usepackage{enumerate}
\usepackage{amsmath}
\usepackage{amsthm}
\usepackage{amssymb}
\usepackage{graphicx}
\usepackage{floatflt}
\usepackage{fancybox}
\usepackage{stmaryrd}
\usepackage{appendix} 
\usepackage{enumitem} 
\usepackage{color}
\usepackage{braket}

\usepackage{hyperref} 
\usepackage[b]{esvect}

\renewcommand{\Re}{\textrm{Re}}
\renewcommand{\Im}{\textrm{Im}}
\newcommand{\Ai}{\textrm{Ai}}

\newcommand{\om}{\omega}
\newcommand{\Om}{\Omega}
\newcommand{\lam}{\lambda}
\newcommand{\eps}{\epsilon}

\newcommand{\sign}{\textrm{sign}}

\newcommand{\p}{\partial}

\newcommand{\m}{{\rm min}}

\newcommand{\be} {\begin{equation}}
\newcommand{\ee} {\end{equation}}
\newcommand{\bsub}{\begin{subequations}}
\newcommand{\esub}{\end{subequations}}
\newcommand{\bea}{\begin{eqnarray}}
\newcommand{\eea}{\end{eqnarray}}
\newcommand{\bi} {\begin{itemize}}
\newcommand{\ei} {\end{itemize}}
\newcommand{\ben} {\begin{enumerate}}
\newcommand{\een} {\end{enumerate}}
\newcommand{\bmat} {\begin{pmatrix}}
\newcommand{\emat} {\end{pmatrix}} 
\newcommand{\bal} {\begin{aligned}}
\newcommand{\eal} {\end{aligned}}
\newcommand{\btab}{\begin{tabular}}
\newcommand{\etab}{\end{tabular}}

\newcommand{\D}{\mathrm{d}}

\begin{document}
\selectlanguage{english}

\title{Quasi-normal modes and fermionic vacuum decay\\ around a Kerr black hole}

\author{Antonin Coutant}
\email{antonin.coutant@ens-lyon.org}
\affiliation{School of Mathematical Sciences, University of Nottingham, University Park, Nottingham, NG7 2RD, United Kingdom}

\author{Peter Millington}
\email{p.millington@nottingham.ac.uk}
\affiliation{School of Physics and Astronomy, University of Nottingham, University Park, Nottingham, NG7 2RD, United Kingdom}

\begin{abstract}
We analyze the instability of the non-rotating fermion vacuum in Kerr spacetimes. We describe how the co-rotating Fermi sea is formed as a result of a spontaneous vacuum decay. Most significantly, and drawing upon intuition gained from analogous electrodynamic processes in supercritical fields, we show that this decay process is encoded entirely in a subset of quasi-normal fermion modes.
\end{abstract}


\maketitle


\section{Introduction}

Nothing can escape from the event horizon of a black hole, not even light, and for that reason, black holes are often thought of as perfect absorbers. Despite this, around rotating black holes, bosonic fields can extract angular momentum from the black hole and subsequently see their energy amplified. For this to occur, their angular frequency $\om$ must satisfy $\om < m \Om_H$, where $m$ is their azimuthal number and $\Om_H$ is the angular velocity of the black hole. This process, known as `superradiance', has been  studied thoroughly since its discovery (see Ref.~\cite{Brito15} for a recent review) and has even been tested experimentally in a water-wave black-hole analogue~\cite{Torres16}. If the black hole is surrounded by a perfect mirror, this amplification process is turned into a dynamical instability: waves propagate back and forth from the mirror to the black hole, amplifying in energy each time and leading to an exponential blow up called the `black hole bomb'. Nature ``can also provide its own mirror''~\cite{Press72}: when the field is endowed with a non-zero mass, low-frequency modes are trapped, thereby triggering a black hole bomb instability~\cite{Cardoso04}. 

On the other hand, fermionic fields around rotating black holes cannot be amplified~\cite{Martellini77,Iyer78} due to the Pauli exclusion principle. However, for a massless fermion, modes satisfying the condition $\om < m\Om_H$ display a vacuum instability: the ergoregion spontaneously emits a steady flux of particles in that frequency range. This is known as the Unruh-Starobinski radiation~\cite{Starobinskii73,Unruh74}. This dual behavior of superradiance is reminiscent of the Klein paradox~\cite{Manogue88}, where amplification occurs only for bosons, but both bosons and fermions can experience a spontaneous emission in strong external fields. When considering fermions with a non-zero mass $\mu$ around a rotating black hole, it has been argued that modes in the range $\om < m\Om_H$ and $\om < \mu$ form a Fermi sea outside the black hole~\cite{Hartman09}. This Fermi sea carries a non-zero angular momentum vacuum expectation value, aligned with the one of the black hole. 

In this work, we show that the Fermi sea is the result of a spontaneous vacuum decay, which constitutes the fermionic pendant of the black hole bomb instability. Indeed, shortly after a rotating black hole is formed, the fermion field is in a vacuum state resembling the Unruh vacuum and carrying zero angular momentum. Subsequently, the Unruh-Starobinski radiation sets in and fills a set of trapped modes outside the black hole until Pauli's exclusion principle stops the process. Through this mechanism, the initial, non-rotating vacuum decays into a new vacuum (the Fermi sea), which co-rotates with the black hole. As we will show, this decay process is entirely encoded in the set of quasi-normal modes~\footnote{Here, we refer to all modes whose frequencies are the poles of the (analytically continued) Green's function as `quasi-normal'~\cite{Berti09,Konoplya11,Zworski17}.} (QNM) that satisfy both conditions $\om < m\Om_H$ and $\om < \mu$. The QNM spectrum of massive Dirac fields has been studied in the literature~\cite{Ternov78,Ternov88,Konoplya07,Dolan09,Konoplya17}, but so far only at the classical level. Notice that the structure of the vacuum states for massless fermions has been studied by various authors~\cite{Casals12,Winstanley13}, and there is no vacuum decay in this case. 

The paper is organized as follows. In section II, we quickly review the Dirac equation in the Kerr spacetime. We then use a WKB approximation to discuss the structure of eigenmode solutions and the set of QNMs in the range $\om < m\Om_H$ and $\om < \mu$. In section III, we present a toy model for the vacuum decay instability, comprising a charged fermion in a strong electric field and tailored to resemble the Kerr problem. In section IV, we discuss the dynamics of the vacuum decay in detail, as well as the main properties of the initial and final states. Throughout the paper, we work in units such that $c = \hbar = G = 1$.


\section{Fermions in Kerr}
\label{Kerr_Classical_Sec}

We consider a rotating black hole, whose spacetime is described by the Kerr metric. Using the Boyer-Lindquist coordinate system, the line element reads 
\bea
\D s^2 \ &=&\ -\left(1 - \frac{2Mr}{\rho^2}\right) \D t^2\: -\: \frac{4aMr \sin^2 \theta}{\rho^2}\, \D t\, \D\phi\: +\:\frac{\rho^2}{\Delta}\, \D r^2 \nonumber\\&+&  \rho^2\, \D\theta^2\: +\: \left(r^2 + a^2 + \frac{2a^2 Mr \sin^2 \theta}{\rho^2} \right) \sin^2 \theta\,\D\phi^2\;, 
\eea
where $\rho^2 = r^2 + a^2 \cos^2 \theta$ and $\Delta = r^2 - 2Mr + a^2$. This spacetime possesses many interesting features and has been studied extensively in the literature~\cite{Chandrasekhar,Straumann,Wiltshire}. In particular, there is an event horizon at $r = r_+ = M + \sqrt{M^2 - a^2}$ (and a Cauchy horizon at $r = r_- = M - \sqrt{M^2 - a^2}$, although the latter won't play any role in what follows). In addition, the asymptotically time-like Killing field $\p_t$ becomes space-like close to the black hole, when $2Mr > \rho^2$. This defines the ergoregion, where any time-like trajectory must co-rotate with the black hole. To characterize the rotation of the black hole, we define its angular velocity 
\be
\Om_H \ =\ \frac{a}{2 M r_+}\;. 
\ee
As we shall see, $\Om_H$ plays an important role in characterizing the vacuum instability that we will analyze. In addition, it is quite convenient to introduce the tortoise coordinate $r_*$, adapted to scattering problems from the horizon to infinity. It is defined via 
\be \label{tort_coord}
\D r_* \ =\  \frac{r^2 + a^2}{\Delta}\,\D r\;. 
\ee
We see that $r_*$ goes from $-\infty$ to $+\infty$ when $r$ runs from $r_+$ to $+\infty$. 

\subsection{The Dirac equation in Kerr}

We now consider non-interacting fermions of mass $\mu$ in this spacetime. These are described by a Dirac field $\psi$, which obeys the equation 
\be \label{D_eq}
(i\gamma^\nu D_\nu \:-\: \mu) \psi\ =\ 0\;, 
\ee
where $D_\nu$ is the spinor covariant derivative. It turns out that, in the Kerr metric, both the massive and massless Dirac equations admit a complete separation of variables. This was first shown by Unruh~\cite{Unruh73} for the massless case and later by Chandrasekhar for the massive case~\cite{Chandrasekhar76,Chandrasekhar}. In this paper, we shall use the same conventions as in Ref.~\cite{Dolan15}; namely, we shall work in the Weyl/Chiral representation, where the flat-spacetime gamma matrices are
\begin{equation}
\hat{\gamma}^0 \ =\ \bmat 0 & I_2 \\ I_2 & 0 \emat\qquad \text{and}\qquad \hat{\gamma}^{j}\ = \ \bmat 0 & \sigma^{j} \\ -\sigma^{j} & 0\emat,
\end{equation}
in which $I_2={\rm diag}(1,1)$ is the two-dimensional unit matrix and the $\sigma^{j}$ ($j =1,2,3$) are the Pauli matrices. We refer the reader to that work~\cite{Dolan15} for a detailed derivation of the Dirac equation and its separation, or our Appendix~\ref{Dirac_CST_App}, where we summarize the various conventions used here.

Due to the symmetries of the metric, the eigenmodes of the Dirac equation \eqref{D_eq} can be written in the form 
\be \label{separated_form}
\psi\ = \ \frac{e^{- i \om t + i m \phi}}{\Delta^{1/4}} \bmat -R_2(r) S_1(\theta)/\sqrt{z} \\ -R_1(r) S_2(\theta)/\sqrt{z} \\ R_1(r) S_1(\theta)/\sqrt{z^*} \\ R_2(r) S_2(\theta)/\sqrt{z^*} \emat\;, 
\ee
where the complex variable $z = r + i a \cos \theta$. In addition, $\om$ is the frequency, and $m$ is a half integer giving the azimuthal number. The above \emph{ansatz} leads to an ordinary differential equation for both the radial and angular parts. First, the angular parts obey the coupled equations 
\bsub \label{angular_mode_eq} \bea
\left( \p_\theta + \frac12 \cot \theta - m \csc \theta + a \om \sin \theta \right) S_1\ &=&\ \left(+\lam_{j, m, \mathcal P}^{(\om, \mu)} + a \mu \cos \theta \right) S_2\;, \\
\left( \p_\theta + \frac12 \cot \theta + m \csc \theta - a \om \sin \theta \right) S_2\ &=&\ \left(-\lam_{j, m, \mathcal P}^{(\om, \mu)} + a \mu \cos \theta \right) S_1\;,
\eea \esub
where $\lam_{j, m, \mathcal P}^{(\om, \mu)}$ is the separation constant, which is closely related to the total angular momentum $j \in \mathbb N + 1/2$ (see Ref.~\cite{Dolan09} for a detailed discussion of the angular equation). In the non-rotating limit, i.e.~when $a \to 0$, $\lam_\om$ reduces to $\mathcal P (j+1/2)$, where $\mathcal P = \pm 1$ is the parity, which we recognize in terms of the eigenvalues of the spin-orbit coupling. However, in this work, we are instead interested in sufficiently rapidly rotating black holes, and $\lam_{j, m, \mathcal P}^{(\om, \mu)}$ is, in general, a complicated function of $j$ and $\mathcal P$, as well as $a \om$ and $a \mu$. As we shall see, the modes responsible for the vacuum decay satisfy $0 < \om \lesssim \mu$. It turns out that the angular equations for $\om = \mu$ can be solved exactly~\cite{Dolan09}, and one can use these solutions to obtain leading-order expressions. In this regime, the separation constant is given by 
\be \label{near_mu_lam}
\lambda_{j, m, \mathcal P}^{(\omega, \mu)} \ =\ -\:\frac12 \:+\: \mathcal P \sqrt{(j + (1+\mathcal P)/2)^2 - 2m a \mu + a^2 \mu^2}\;. 
\ee
To simplify our notation, we hereafter drop the indices on $\lam^{(\omega,\mu)}_{j,m,\mathcal{P}}\equiv \lam$. Once $\lam$ is determined, the radial part obeys the coupled equations 
\bsub \label{rad_mode_eq} \bea
\sqrt{\Delta} \left( \p_r - i\frac{K_\om}{\Delta} \right) R_1\ &=&\ \left( \lam + i \mu r \right) R_2\;, \\
\sqrt{\Delta} \left( \p_r + i\frac{K_\om}{\Delta} \right) R_2\ &=&\ \left( \lam - i \mu r \right) R_1\;, 
\eea \esub
with 
\be
K_\om\ =\ (r^2+a^2) \om\: -\: a m\;. 
\ee
The aim now is to solve equation \eqref{rad_mode_eq} with appropriate boundary conditions.


\subsection{Complete mode basis}

Once the mode equations \eqref{angular_mode_eq} and \eqref{rad_mode_eq} have been solved, general solutions of the Dirac equation \eqref{D_eq} are obtained by linear superpositions. In particular, the field operator decomposes into a sum of eigenmodes, where the associated operator-valued amplitudes are the familiar creation and annihilation operators. Formally, the decomposition has the form 
 \be
\hat \psi(t,r,\theta,\phi)\ =\ \sum_{j, m, \mathcal P, \om} \left( \hat a^{(j, m, \mathcal P)}_\om \varphi^{(j, m, \mathcal P)}_\om\: +\: (\hat b^{(j, m, \mathcal P)}_\om)^\dagger \psi^{(j, m, \mathcal P)}_\om \right)\;. 
\ee
Once $(j, m, \mathcal P)$ is chosen, the modes are determined by solutions of equation \eqref{rad_mode_eq} with specific boundary conditions. (In the following, we drop the superscript $(j, m, \mathcal P)$, as we have done for the eigenvalue $\lambda$.) There are now two families of modes. The first one is defined at infinity ($r\to \infty$) as incoming towards the black hole. These modes are asymptotically on-shell and hence exist only above the mass gap $\om > \mu$. Following standard conventions, we refer to them as \emph{in}-modes. Moreover, we have
\be
\psi_\om^{\rm in} \ =\ \varphi_{-\om}^{\rm in}\;. 
\ee
The second family is defined near the horizon ($r \to r_+$) as emanating from it, and we refer to these as \emph{up}-modes. The positive and negative continua are related by 
\be
\psi_\om^{\rm up} \ =\ \varphi_{2m\Om_H - \om}^{\rm up}\;. 
\ee
Therefore, the field operator has the general eigenmode decomposition 
\be
\hat \psi(t,r,\theta,\phi)\ =\ \sum_{j, m, \mathcal P}\left[ \int_{\om > \mu}  \D\omega \left( \hat a_\om^{\rm in} \varphi_\om^{\rm in} + (\hat b_\om^{\rm in})^\dagger \psi_\om^{\rm in} \right)\:  +\:\ \int_{\om > m \Om_H}\D\omega \left( \hat a_\om^{\rm up} \varphi_\om^{\rm up} + (\hat b_\om^{\rm up})^\dagger \psi_\om^{\rm up} \right) \right]\;, 
\ee
where $\hat a_\om$ and $\hat b_\om$ satisfy the canonical anti-commutation relations. These anti-commutation relations are guaranteed when using normalized modes with respect to the (conserved) scalar product, given by 
\be
\langle \psi_1 | \psi_2 \rangle \ =\  \int \D r\, \D\Om\;\rho^2 \bar \psi_1 \gamma^0 \psi_2\;. 
\ee
Using the \emph{ansatz} \eqref{separated_form}, the norm of a mode separates into 
\be
\label{eq:sep}
\langle \psi | \psi \rangle \ =\ \int_{r_+}^{\infty}\D r\left(|R_1|^2\: +\: |R_2|^2 \right) \frac{r^2+a^2}{\Delta}\:  \times\: 2\pi \int_{-1}^1 \D\cos(\theta)\left(|S_1|^2\: +\: |S_2|^2 \right)\;. 
\ee
Notice that the radial part takes the canonical form $\int_{\mathbb R}\D r_* \left(|R_1|^2 + |R_2|^2 \right)  $ using the tortoise coordinate \eqref{tort_coord}. Since the stationary modes form a dense set, we must use a Dirac normalization, that is, we require 
\bsub \label{modes_norm} \bea
2\pi \int_{-1}^1 \D\cos(\theta) \left(|S_1|^2\: +\: |S_2|^2 \right) \  &=&\ 1\;, \\ 
\int_{\mathbb R} \D r_*\left( R_{1, \om}^* R_{1, \om'}^{\phantom{*}} \:+\: R_{2, \om}^* R_{2, \om'}^{\phantom{*}} \right) \ &=&\ \delta(\om - \om')\;,  \label{modes_norm_rad}
\eea \esub
with a vanishing overlap for differing values of $(j, m, \mathcal P)$. As usual, when dealing with a dense set of modes, it is simpler to normalize the conserved current. Hence, for the radial part, we require 
\be \label{current_norm}
\oint \D\Om\;\rho^2 J^r \ =\ |R_1|^2\: -\: |R_2|^2\ =\ \pm\,\frac1{2\pi}\;. 
\ee
As we show in detail in appendix~\ref{Norm_App}, the conservation law~\cite{Dolan15}
\be \label{Kerr_conservation}
\p_t \oint \D\Om\;\rho^2 J^t  \:+\: \p_r \oint \D\Om\;\rho^2 J^r\ =\ 0\; ,
\ee
with $\oint \D\Om\,\rho^2 J^t  = \int_{\mathbb R}\D r_* \left(|R_1|^2 + |R_2|^2 \right)$, guarantees that the current normalization \eqref{current_norm} is equivalent to the norm \eqref{modes_norm_rad}.


\subsection{Eigenmodes and QNMs in the WKB approximation}
\label{QNM_Sec}
\begin{figure}[!ht]
\begin{center}
\includegraphics[width=0.49\textwidth,trim=0 0 5mm 0]{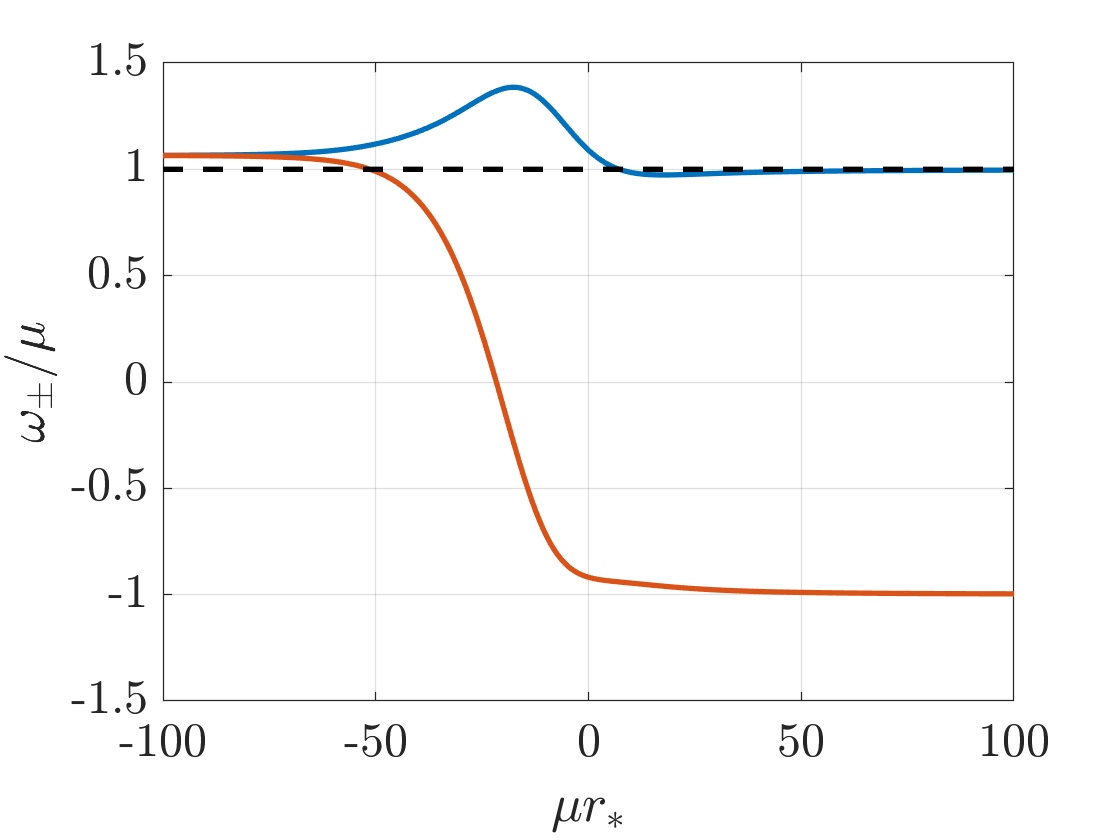}\includegraphics[width=0.49\textwidth,trim=0 0 5mm 0]{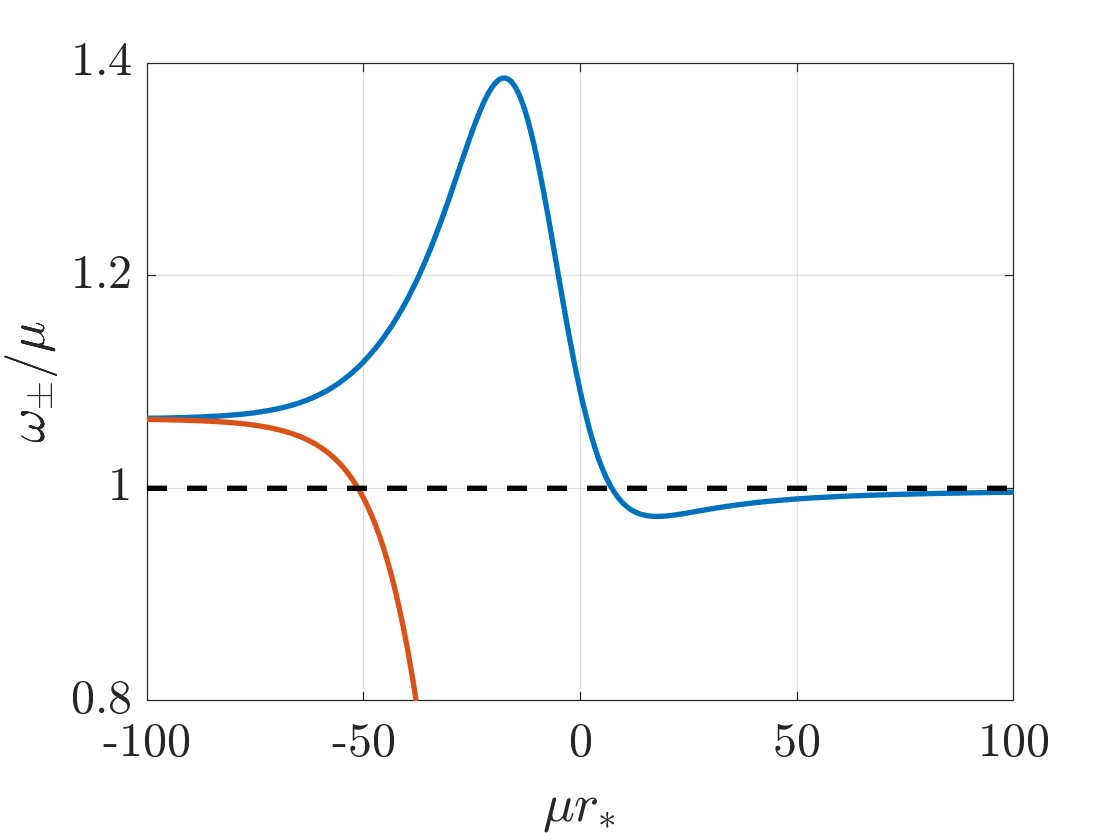}
\end{center}
\caption{Left side: $\om_+$ (blue) and $\om_-$ (red) as a function of $r_*$ for $0 < \om < \min(\mu, m\Om_H)$. The parameters are chosen such that $M \mu = 0.22$, $a/M=0.999$, $\om/\mu = 0.99$, $m=1/2$, and $\lam$ is given by equation~\eqref{near_mu_lam} with $(j,\mathcal P) = (1/2,1)$. The line of constant $\om$ (dashed line) crosses $\om_\pm$ three times: $\om_+$ twice at $r_2$ and $r_3$, which forms the trapping region, and $\om_-$ once at $r_1$. Right side: magnified view of the trapping region.
}
\label{potential_Fig} 
\end{figure}

To understand the structure of the eigenmode solutions of equation \eqref{rad_mode_eq}, we shall employ a WKB method. The WKB approximation is particularly good in the regime $M \mu \gg 1$~\cite{Dolan15}. Since, from now on, we shall mainly focus on the radial part, we use the condensed notation 
\be
R(r) \ =\ \bmat R_1(r) \\ R_2(r) \emat\;. 
\ee
To proceed with the analysis of equation \eqref{rad_mode_eq}, we assume 
\be \label{WKB_ansatz}
R(r)\ =\ \bmat R_1^0(r) \\ R_2^0(r) \emat e^{i \int {\rm d}r\; k(r) }
\ee
and that the amplitudes $R_j^0$ vary much more slowly than the phase $\int {\rm d}r\; k(r)$. This leads to the $2 \times 2$ linear system 
\bsub \label{rad_Ampl_WKB} \bea
\sqrt{\Delta} \left( k - \frac{K_\om}{\Delta} \right) R_1^0\: +\: i \left( \lam + i \mu r \right) R_2^0 \ &=&\ 0\;, \\
\sqrt{\Delta} \left( k + \frac{K_\om}{\Delta} \right) R_2^0 \:+\: i \left( \lam - i \mu r \right) R_1^0 \ &=&\ 0\;.  
\eea \esub
This system admits non-trivial solutions if the determinant of the coefficient matrix vanishes. This gives the dispersion relation or, equivalently, the Hamilton-Jacobi equation 
\be \label{HJ_eq}
k^2 \ =\ \frac{K_\om^2}{\Delta^2} \: -\: \frac{\lam^2 + \mu^2 r^2}{\Delta}\;. 
\ee
Moreover, equation \eqref{rad_Ampl_WKB} gives a linear relation between $R_1^0$ and $R_2^0$. The additional requirement of unit current \eqref{current_norm} leads to the full expression of the amplitudes for a given $k$; namely, 
\be \label{WKB_Ampl}
R_1^0 \ =\ \frac{i \lam - \mu r}{\sqrt{4\pi |k| (K_\om - \Delta k)}} \qquad \text{and} \qquad R_2^0 \ =\ -\sqrt{\frac{K_\om - \Delta k}{4\pi |k| \Delta}}\;. 
\ee
As we see in the Hamilton-Jacobi equation \eqref{HJ_eq}, depending on the sign of the right-hand side, the modes are either propagating or evanescent. A convenient way to discuss its sign is to recast equation \eqref{HJ_eq} in the form~\cite{Damour76b,Damour78}  
\be
k^2 \ =\ \frac{(r^2+a^2)^2}{\Delta^2}\,\big(\om - \om_+(r)\big)\big(\om - \om_-(r)\big)\;. 
\ee
Around a certain radius $r$, the modes propagate if $\om > \om_+$ or if $\om < \om_-$. Hence, $\om_\pm$ conveniently delimitate the positive- and negative-frequency continua. The explicit expressions for $\om_\pm$ follow from the Hamilton-Jacobi equation \eqref{HJ_eq}, and we have   
\be
\om_\pm(r)\ =\ \frac{a m \pm \sqrt{\Delta (\lam^2 + \mu^2 r^2)}}{r^2+a^2}\;. 
\ee
In Fig.~\ref{potential_Fig}, we have represented $\om_\pm$ as functions of $r$. We first notice the behavior of $\om_\pm$ on both asymptotic regions (i.e.~near the horizon and at infinity): 
\bsub \bea
\om_\pm(r) \ &\underset{\infty}{\sim}&\ \pm \mu\;, \\
&\underset{r_+}{\sim}&\ m \Om_H\;.
\eea \esub

A remarkable fact of rotation is that frequencies satisfying $0 < \om < m \Om_H$ can tunnel from the lower continuum close to the horizon to the upper continuum outside the potential barrier. If such a frequency is also above the mass gap $\om > \mu$, it tunnels out and propagates to infinity. This gives rise to the Unruh-Starobinski radiation. In this work, we are interested in the frequencies below the mass gap that are also able to tunnel to another propagating region (such that $\om > \om_+$). This happens if $\om$ is above the local minimum of $\om_+$ (see Fig.~\ref{potential_Fig}), which we call $\om_\m$. These frequencies therefore satisfy $\om_\m < \om < \mu$ and $\om < m \Om_H$. In this case, there are three turning points $r_{j}$ ($j=1, 2, 3$) such that $\om_\pm(r_{j}) = \om$, and the modes are propagating for $r < r_1$ and $r_2 < r < r_3$. On the horizon side, for $r < r_1$, the modes are superpositions of right movers and left movers with a scattering phase shift $e^{2i \delta_\om}$~\cite{Newton}, i.e. 
\be \label{Kerr_modes_left}
R(r) \ \simeq\ 
\bmat \frac{i \lam - \mu r}{\sqrt{4\pi |k| (K_\om - \Delta k)}} \\ -\sqrt{\frac{K_\om - \Delta k}{4\pi |k| \Delta}} \emat e^{+i \int_{r_1}^r \D r'k(r') } \:+\: \bmat \frac{i \lam - \mu r}{\sqrt{4\pi |k| (K_\om + \Delta k)}} \\ -\sqrt{\frac{K_\om + \Delta k}{4\pi |k| \Delta}} \emat e^{-i \int_{r_1}^r\D r' k(r')  + 2i \delta_\om}\;. 
\ee
Notice that, at this point and onward, $k$ denotes only the positive root of the Hamilton-Jacobi equation \eqref{HJ_eq}. For $r_2 < r < r_3$, the modes are again a superposition of left and right movers, with an overall amplitude $A_\om$, which depends on the tunneling probability. Explicitly, we have 
\be \label{Kerr_modes_right}
R(r)\ \simeq\ A_\om \left[\bmat \frac{i \lam - \mu r}{\sqrt{4\pi |k| (K_\om - \Delta k)}} \\ -\sqrt{\frac{K_\om - \Delta k}{4\pi |k| \Delta}} \emat e^{+i \int_{r_3}^r  \D r' k(r') - i \frac\pi4} \:+\: \bmat \frac{i \lam - \mu r}{\sqrt{4\pi |k| (K_\om + \Delta k)}} \\ -\sqrt{\frac{K_\om + \Delta k}{4\pi |k| \Delta}} \emat e^{-i \int_{r_3}^r \D r' k(r')  + i \frac\pi4}\right]\;. 
\ee
Using connection formula on the turning points~\cite{Berry72,Gottfried} (see also appendix \ref{app:turning}), we obtain both the scattering phase shift and the amplitude in the trapped region:
\bsub \bea
e^{2i \delta_\om}\ &=&\ \frac{A_\om^*}{A_\om}\;, \\
A_\om \ &=&\ \frac{e^{-I_\om + i \frac{\pi}4}}{\cos(S_\om) - i \sin(S_\om) e^{-2 I_\om}}\;. \label{A_eq}
\eea \esub
In these equations, $I_\om$ and $S_\om$ are the classical actions in the forbidden and allowed regions, that is 
\bsub \bea
S_\om \ &=&\ \int_{r_2}^{r_3} \D r\; k(r)\;, \\
I_\om \ &=&\  \int_{r_1}^{r_2} \D r\; |k(r)|\;.
\eea \esub

The expression \eqref{A_eq} for $A_\om$ shows a set of resonant frequencies near $\cos(S_\om) = 0$. These resonances are poles of $A_\om$ in the complex plane. Since they are also poles of the phase shift $e^{2i\delta_\om}$, we see that they are associated with in-going boundary conditions at the horizon (see equation~\eqref{Kerr_modes_left}), and since they are out-going at spatial infinity by construction, these poles are quasi-normal frequencies. Since they are trapped due to the condition $\om < \mu$, they are usually called quasi-bound states (QBS)~\cite{Dolan15} and are non-radiative, contrary to usual QNMs. It is convenient to write the quasi-normal frequencies $\mathfrak{w}$ as a sum of real and imaginary parts, viz.~$\mathfrak{w} = \om^R + i \om^I$. Using the WKB analysis outlined above, and in the limit of small tunneling $|\om^I | \ll |\om^R|$, we obtain an approximate expression for the quasi-normal frequencies. The real part is obtained by solving a Bohr-Sommerfeld condition, and the imaginary part follows from the tunneling amplitude: 
\bsub \label{eq:QNMdef}\bea
S_{\om_n^R} \ &=&\ \pi (n+1/2)\;, \label{QNM_Re}\\
\om^I_n \ &=&\ -\: \dfrac{\exp\left(-2 I_{\om_n^R}\right)}{T_{\rm cl}(\om_n^R)}\;, \label{QNM_Im}
\eea \esub
where $n$ is an integer and
\be
\label{eq:Tcl}
T_{\rm cl} \ =\ \p_\om S\ =\ \int_{r_2}^{r_3}\frac{{\rm d}r'}{v_g(r')}
\ee
is the time needed for a semiclassical wave packet to propagate at a local group velocity $v_g(r) = 1/\p_\om k(r)$ from $r_2$ to $r_3$. As we shall see, this particular set of QNMs encodes a vacuum instability.

\section{Electric toy model}
\label{sec:electrictoy}

To illustrate the mechanism of the vacuum decay in Kerr, we shall first consider a simple toy model in an electric field. Analogous toy models have been used in the past to study the black hole bomb of scalar fields, see, e.g., the appendix of Ref.~\cite{Arvanitaki09}. Our model comprises a Dirac fermion in $1+1$ dimensions, which is subject to an external electric potential $V(x)$ and has a local mass $\mu(x)$. The two functions will be chosen to mimic the Kerr problem, while yielding much simpler calculations. Moreover, this electric model allows us to make connection with the literature on fermions in overcritical fields~\cite{Greiner,Reinhardt:1977ps,Rafelski78}, which bare many similarities with black hole physics~\cite{Primer}.

The Dirac equation for the mode of energy $\omega$ is
\be \label{Toy_rad_eq} 
\om \psi_\om(x) \: +\: i \gamma^0 \gamma^1 \p_x \psi_\om(x)\: -\: V(x) \psi_\om(x)\: -\: \mu(x) \gamma^0 \psi_\om(x)\ = \ 0\;.
\ee
Note that the coordinate $x$ here is the analogue of $r_*$ in the BH case. As in previous sections, we work in the Weyl/Chiral representation. This means, in particular, that 
\be
\gamma^0 \ =\ \bmat 0 & I_2 \\ I_2 & 0 \emat \qquad \textrm{and} \qquad \gamma^0 \gamma^1 \ =\ \bmat -\,\sigma^{3} & 0 \\ 0 & \sigma^{3} \emat\;,
\ee
where $I_2={\rm diag}(1,1)$ is the two-dimensional unit matrix and $\sigma^{3}=\mathrm{diag}(1,-\,1)$ is the third Pauli matrix.

We assume that $V(x)$ and $\mu(x)$ are piece-wise constant. Alternatively, one could work with slowly varying, smooth $V(x)$ and $\mu(x)$, and use a WKB approximation, as employed for the Kerr problem in Sec.~\ref{Kerr_Classical_Sec}. As we shall see, the piece-wise constant problem is already quite similar to the Kerr problem, without generating extra technicalities. In each region where both $V$ and $\mu$ are constant, the mode solutions of equation~\eqref{Toy_rad_eq} have the form 
\be
\psi_\om(x)\ =\ \psi_{\omega}(0) e^{-ikx}\;,
\ee
where $k$ is the wavenumber. The dispersion relation can be found by acting on equation~\eqref{Toy_rad_eq} from the left with the conjugate Dirac operator. This gives 
\be
(\om \:-\: V)^2 \ =\ k^2\: +\: \mu^2\;.
\ee

\begin{figure}[!ht]
\begin{center}
\includegraphics[width=0.49\textwidth]{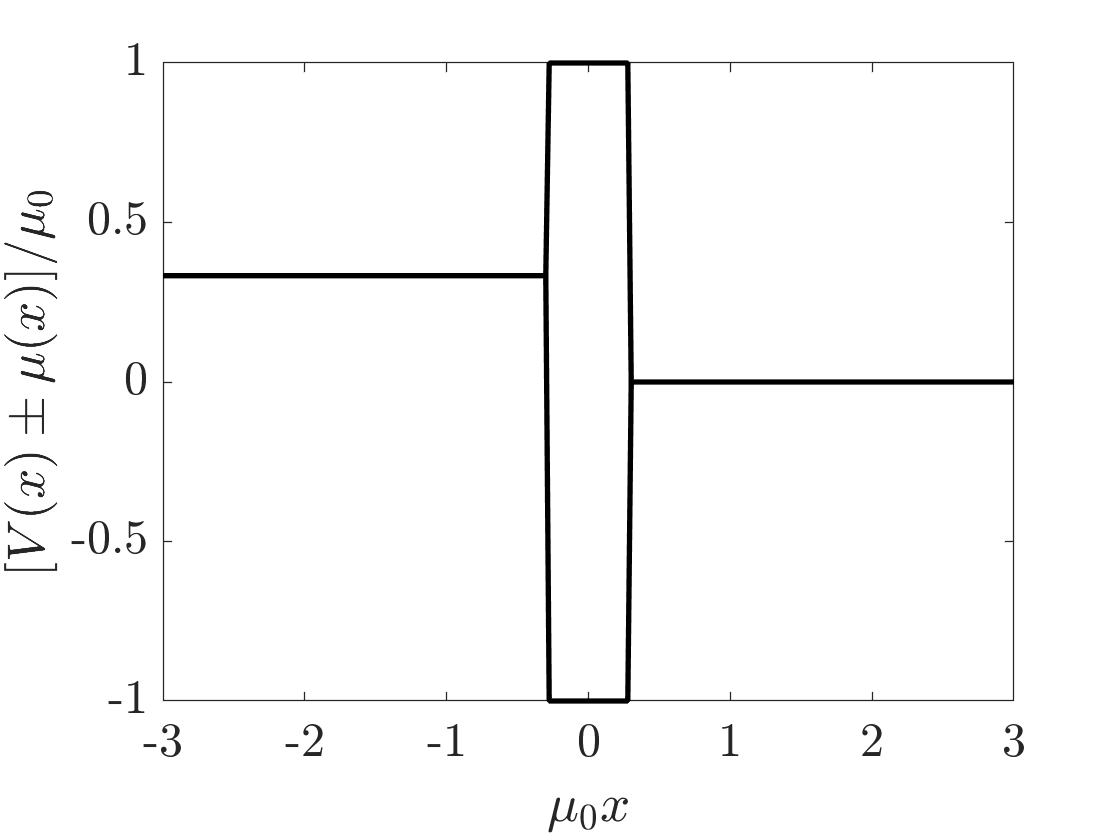}
\end{center}
\caption{Plot of $V(x)\pm \mu(x)$, with $V_0 = 1$, $\mu_0 = 3$ and $D=0.1$. To be compared with Fig.~\ref{potential_Fig}. 
}
\label{Elec_potential_Fig} 
\end{figure}

We now consider a problem with two main regions in which the field is massless. For $x<0$, the potential takes the constant value $V_0$, and it vanishes ($V=0$) for $x>0$. Moreover, we assume that there is a perfectly reflecting boundary at $x=L$. $V_0$ is the electric analogue of $m\Omega_H$ in the Kerr case, and the reflecting condition at $x=L$ mimics modes in Kerr below the mass gap ($\om < \mu$), which are trapped in a finite region near the black hole. In addition, we add a delta-barrier potential at the transition, i.e.~near $x=0$. It is, however, delicate to define a delta-barrier potential for fermions, due to the first-order character of the Dirac equation~\cite{Roy93}. To obtain such a barrier, we assume that there is a small region $-D < x < D$ within which the field has a large mass $\mu_0$. We then take the width of the barrier to zero and the height to infinity such that the product 
\be \label{Fermion_delta}
\xi \ =\ 2\mu_0 D
\ee
remains constant. The parameter $\xi$ allows us to control the `strength' of the barrier: in the limit $\xi \to 0$, there is complete transmission; in the limit $\xi\to \infty$, the barrier is impermeable. Note also that the perfectly reflecting boundary condition at $x=L$ is obtained with the same method by taking $\xi \to \infty$. The effective dispersion relation of the problem is represented in Fig.~\ref{Elec_potential_Fig}. 

The Dirac mode equation \eqref{Toy_rad_eq} has four linearly independent solutions: comprising both spin projections of the positive- and negative-frequency branches. Applying the thin-barrier limit described above, we obtain the boundary condition at $x=0$. This boundary condition relates the value of the mode on each side. Explicitly, we have 
\be \label{BC_elec1}
\psi_\om(0^-) \ =\ \bmat \cosh(\xi) & 0 & -\,i \sinh(\xi) & 0 \\ 
0 & \cosh(\xi) & 0 & i \sinh(\xi) \\
i \sinh(\xi) & 0 & \cosh(\xi) & 0 \\
0 & -\,i \sinh(\xi) & 0 & \cosh(\xi) \\
\emat \cdot \psi_\om(0^+)\;.
\ee
Taking the limit $\xi\to \infty$ of this expression also yields the reflecting boundary condition at $x=L$:
\be \label{BC_elec2}
\bmat 1 & 0 & i & 0 \\ 
0 & 1 & 0 & -\,i \\
-\,i & 0 & 1 & 0 \\
0 & i & 0 & 1 \\
\emat \cdot \psi_\om(L) \ =\ 0\;. 
\ee
Notice that the boundary conditions mix the two chiralities, which reside in the upper and lower pairs of elements in the four-component spinor. Recalling that the spin operator is $\Sigma = \mathrm{diag}(\sigma^3,\sigma^3)$, we see that the energy eigenstates are also spin eigenstates, with the two spin polarizations evolving independently.

We now consider the setup where the central barrier is initially infinite, i.e.~$\xi = \infty$ for $t<0$, and is suddenly lowered to a finite (but large) value for $t>0$. Initially, the eigenmodes compose a continuous set on the left side, reflected on the impermeable barrier, and a discrete set between $0 < x < L$. We denote by $(\varphi_\om^{\rm in})_{\om > V_0}$ (resp.~$(\psi_\om^{\rm in})_{\om > V_0}$) the continuous modes of positive (resp.~negative) energy, and $(\varphi_n^{\rm in})_{n \in \mathbb N_0}$ (resp.~$(\psi_n^{\rm in})_{n \in \mathbb N_0}$) the discrete ones of positive energy (resp. negative). Since the barrier is infinite, these two sets are uncoupled. When the barrier is lowered, the two regions become coupled. In particular, discrete modes of positive energy with $\om_n < V_0$ are coupled to the continuous set of negative energy. As a consequence, the initial vacuum will spontaneously populate these discrete states with particles by emitting anti-particles to the left. This process is the decay of the initial neutral vacuum into the true ground state: a charged vacuum. After the barrier is lowered, for $t>0$, the eigenmodes compose only a continuous set extending over $x \in ]-\infty; L]$ and denoted $(\varphi_\om^{\rm in})_{\om > V_0}$ for positive energies and $(\psi_\om^{\rm in})_{\om > V_0}$ for negative ones. As we will now show, the decay process described before is encoded in the overlap between this continuous set and the discrete set of initial eigenmodes. 

In Appendix~\ref{Eletric_App}, we present the full set of eigenmode solutions of equation~\eqref{Toy_rad_eq}. Since we are interested only in the overlap between the initial discrete modes and the outgoing continuum modes, we can restrict our attention to the initial discrete modes $\varphi_{n}^{\rm in}$ of the cavity $[0, L]$ and their charge conjugates $\psi_{n}^{\rm in}$, as well as the outgoing continuum modes on the interval $[0,L]$. Moreover, since the two spin eigenstates decouple, it is sufficient to consider only one spin polarization. The discrete set of \emph{in}-modes is given by 
\be
\varphi_{n}^{\rm in}(x) \ =\ \frac1{\sqrt{2L}} \bmat e^{-i \om_n x - i \pi/4} \\ 0 \\ e^{+i \om_n x + i \pi/4} \\ 0 \emat \qquad \textrm{and} \qquad \psi_{n}^{\rm in}(x)\ =\ \frac1{\sqrt{2L}} \bmat e^{+i \om_n x - i \pi/4} \\ 0 \\ e^{-i \om_n x + i \pi/4} \\ 0 \emat\;,
\ee
where
\be
\om_n \ =\ \frac{\pi}{L} \left(n+\frac12 \right)\;, \qquad \textrm{with} \qquad n \ \in\ \mathbb N_0\;.
\ee
The continuous set of outgoing modes has a structure very similar to the one in Kerr [see equations \eqref{Kerr_modes_left} and \eqref{Kerr_modes_right}]. On the left side $(x< 0)$, it is a superposition of left and right movers with a scattering phase shift. However, we are mostly interested in these modes in the trapped region ($0 < x < L$), where they read 
\be \label{Elec_out_modes}
\varphi_{\om}^{\rm out}(x) \ =\ \frac{A_\om}{\sqrt{2L}} \bmat e^{-i \om (x-L) + i \pi/4} \\ 0 \\ e^{+i \om (x-L) - i \pi/4} \\ 0 \emat\qquad \text{and}\qquad  \psi_{\om}^{\rm out}(x)\ =\ \varphi_{2V_0 - \om}^{\rm out}(x)\;,
\ee
with
\be
A_\om\ =\ \sqrt{\frac{L}{\pi}}\,\frac{e^{\xi}}{\sin(\om L) \:+\: i \cos(\om L) e^{2\xi}}\;.
\ee
We can now decompose the field operator for both time regions as
\be \label{InOut_Field_Decomp}
\hat \psi(t, x) \ =\ \begin{cases}\sum_{\om_n\, >\, 0}\Big[ \hat a_{n}^{\rm in} \varphi_n^{\rm in}(x) e^{- i \om_n t}\: +\: (\hat b_{n}^{\rm in})^\dagger \psi_n^{\rm in}(x) e^{+i \om_n t} \Big] \\
\ +\:\int_{\om\, >\, V_0}{\rm d}\om\; \Big[ \hat a_{\om}^{\rm in} \varphi_\om^{\rm in}(x) e^{- i \om t} \:+\: (\hat b_{\om}^{\rm in})^\dagger \psi_\om^{\rm in}(x) e^{+i (\om - 2V_0) t} \Big]\;, &\quad t\ <\ 0\;,\\[1em]
\int_{\om\, >\, V_0} {\rm d}\om\;\Big[\hat a_{\om}^{\rm out} \varphi_{\om}^{\rm out}(x) e^{- i \om t}\: +\: (\hat b_{\om}^{\rm out})^\dagger \psi_{\om}^{\rm out}(x) e^{+i (\om - 2V_0) t} \Big]\;, &\quad t\ >\ 0\;.\end{cases}
\ee
Pre-empting the analysis of the Kerr problem that follows in Sec.~\ref{sec:Kerrvac}, we draw attention to the similarity of the $\varphi^{\rm in}$ and $\varphi^{\rm out}$ modes on the interval $[0,L]$. Most importantly, when $\xi \gg 1$, the amplitude $A_{\omega}$ is maximal for $\omega \sim \omega_n$, such that the main distinction between the $\varphi^{\rm in}$ and $\varphi^{\rm out}$ is in their normalization.

We require the field operator to be continuous at $t=0$. This gives a linear relationship between the two representations of the field operator, otherwise known as a Bogoliubov transformation~\cite{Fulling}. We have, for instance, that
\begin{subequations}
 \bea
\hat b_{\om}^{\rm out} &\underset{V_0\, <\, \om}\approx& \sum_{\omega_n\,>\,0} (\hat a_{n}^{\rm in})^{\dag}\,\langle  \varphi_{n}^{\rm in}| \psi_{\om}^{\rm out} \rangle \:+\: \int_{\om'\, >\, V_0}{\rm d}\om'\; \hat b_{\om'}^{\rm in}\,\langle \psi_{\om'}^{\rm in} | \psi_\om^{\rm out} \rangle\\
&\doteq& \sum_{\omega_n\,>\,0} \beta_n(\om) (\hat a_{n}^{\rm in})^{\dagger}\: +\: \int_{\om'\,>\,V_0}{\rm d}\om'\; \alpha_{\om \om'} \hat b_{\om'}^{\rm in}\;,
 \label{App_b_decomp}
\eea
\end{subequations}
where, for brevity, we have included only the contributions from the nearby modes with which $\psi_{\om}^{\rm out*}(x)$ has significant overlap. The coefficients $\beta_n(\omega)$ and $\alpha_{\omega\omega'}$ describe the Bogoliubov transformation between the two representations. The mode basis is orthonormal with respect to the conserved scalar product of the one-particle Hilbert space, i.e. 
\be
\langle \psi | \psi \rangle \ =\ \int{\rm d}x\; \psi^\dagger(x) \psi(x)\;. 
\ee 
Moreover, as we will see below above (see also appendix~\ref{Eletric_App}), the most relevant overlap for $0 < \om < V_0$ is 
\be \label{Electric_overlap}
\langle \varphi_n^{\rm in} | \varphi_\om^{\rm out} \rangle\ =\ -\,2A_\om\, \frac{\sin[(\om - \om_n)L/2] \sin[(\om + \om_n)L/2]}{(\om - \om_n)L}\;,
\ee
thanks to the identity
\bsub \label{Beta_App} \bea
\beta_n(2V_0 - \om) \ &=&\ \langle \varphi_n^{\rm in} | \psi_{2V_0 - \om}^{\rm out} \rangle \label{Elec_beta} \\
&=& \ \langle \varphi_n^{\rm in} | \varphi_\om^{\rm out} \rangle\;. 
\eea \esub
Notice that equation \eqref{Electric_overlap} is the analytic continuation of \eqref{Elec_out_modes} for $0 < \om < V_0$. Hence, it physically corresponds to an overlap between a positive energy mode with a negative energy one as in equation \eqref{Elec_beta}. 

We now have all the ingredients that we need to understand the decay of the vacuum state. To describe this decay, we first define an instantaneous occupation number for the discrete modes. This will allow us to follow their evolution from initially empty to occupied, at the end of the decay process. The natural definition of the instantaneous mean occupation number of the state $|\varphi_n^{\rm in} \rangle$ is~\cite{Greiner,Reinhardt:1977ps,Rafelski78} 
\be
N_n(t) \ =\ \langle 0^{\rm in}|\hat a_n^\dagger(t) \hat a_n(t) |0^{\rm in} \rangle\;,
\ee
where we have defined the instantaneous annihilation operator of the (one-particle) state $|\varphi_n^{\rm in} \rangle$ 
\be \label{Elec_instant_a}
\hat a_n(t)\ =\ \langle \varphi_n^{\rm in} | \hat \psi \rangle\ =\ \int {\rm d}x\;\varphi_n^\dagger(x) \hat \psi(t,x) \;. 
\ee

For $t<0$, $|\varphi_n^{\rm in}\rangle$ is a stationary state, and $\hat{a}_n(t)$ evolves with the phase $e^{-i \om_n t}$. For $t>0$, $|\varphi_n^{\rm in}\rangle$ is no longer a stationary state, and $\hat{a}_n(t)$ is instead associated with a resonance. As a result, the vacuum state for $t<0$ is no longer stationary for $t>0$, and it decays to the true ground state. If one reinstates the barrier ($\xi \to \infty$) after some time $T$, $\hat a_n(t)$ corresponds again to a stationary state, and $N_n(T)$ will give the mean number of particles created in the state $n$ between $t=0$ and $t=T$. This justifies our definition of the instantaneous occupation number.

We anticipate that  the state $|\varphi_n^{\rm in} \rangle$ will initially be empty for $0 < \om_n < V_0$, subsequently decaying to an occupied state and meaning that $N_n(t) = 1-(\textrm{decaying terms})$. Hence, it is more convenient to use the anti-commutation relations to write
\be
N_n(t) \ =\ 1\: -\: \langle 0^{\rm in}|\hat a_n(t) \hat a_n^\dagger(t)|0^{\rm in} \rangle\:. 
\ee
We now use equation \eqref{Elec_instant_a} to obtain
\bea
\hat a_n^\dagger(t)\ =\ \int_{\omega\,>\,V_0}{\rm d}\om\;\Big[ \langle \varphi_\om^{\rm out} | \varphi_n^{\rm in} \rangle (\hat{a}_\om^{\rm out})^\dagger e^{+i \om t} \:+\:\langle \psi_\om^{\rm out} | \varphi_n^{\rm in} \rangle \hat{b}_\om^{\rm out} e^{- i (\om - 2V_0) t} \Big]\;, 
\eea 
substituting for $b_{\om}^{\rm out}$ from equation~\eqref{App_b_decomp}.
At leading order in $e^{-\xi}$, the overlap $\langle \varphi_n^{\rm in} | \varphi_\om^{\rm out} \rangle$ is negligible on the integration interval, because $\om_n < V_0 < \om$ (see equation~\eqref{Electric_overlap}), and we find
\be
\hat a_n^\dagger(t) |0^{\rm in} \rangle\ =\ \sum_{\omega_n'\,>\,0} \int_{\omega\, >\, V_0} {\rm d} \omega\; \beta_n(\om)^* \beta_{n'}(\om) e^{- i (\om - 2V_0) t} | 1_{n'}^{\rm in} \rangle\;,
\ee
such that
\be
N_n(t)\ =\ 1 \:-\: \sum_{\omega_n'\,>\,0} \left| \int_{\omega\, >\, V_0} {\rm d} \omega\; \beta_n(\om)^* \beta_{n'}(\om) e^{- i (\om - 2V_0) t} \right|^2\;. 
\ee
From equations~\eqref{Electric_overlap} and \eqref{Beta_App}, we see that $\beta_n(\om)$ is proportional to $\mathrm{sinc}[(\om-\om_n)L/2]$. Hence, at leading order in $e^{-\xi}$, $\beta_n(\om)^* \beta_{n'}(\om) \propto \delta_{n n'}$, and we can drop the sum in the above equation. Using the change of variable $\om \to 2V_0 - \om$ and equation~\eqref{Beta_App}, it follows that
\be
N_n(t) \ =\ 1\: -\: \left| \int_{-\infty}^{V_0} {\rm d} \omega\; |\langle \varphi_{n}^{\rm in} | \varphi_{\om}^{\rm out} \rangle|^2 e^{-i \om t} \right|^2\;. 
\ee

All that remains is to compute the integral over $\omega$ by the residue theorem. We see from equation \eqref{Electric_overlap} that all the singularities of the overlap are those of $F(\omega)=|A_{\omega}|^2$. Once extended in the complex plane, the pole and corresponding residue of $F(\om)$ in the lower-half complex plane are, to leading order in $e^{-\xi}$, given by 
\bsub \bea
\mathfrak{w}_n \ &=&\ \om_n\: -\: i\, \frac{e^{-2\xi}}{L}\:, \\
\mathrm{Res}_n(F)\ &=&\ -\:\frac{1}{2 i \pi}\;. 
\eea \esub
For $t>0$, we pick up the poles in the lower-half plane. Moreover, the factor $\mathrm{sinc}[(\om-\om_n)L/2]$ in the overlap selects only the pole $\mathfrak{w}_n$ whose real part $\omega^R_n$ is close to $\om_n$. We therefore obtain 
\bea
\int{\rm d}\om\; |\langle \varphi_{n}^{\rm in} | \varphi_{\om}^{\rm out} \rangle|^2 e^{-i \om t}\ &=&\ -\:2i \pi\: \times\: 4\: \times\: \mathrm{Res}_n(F)\: \times\: e^{- i \mathfrak{w}_n t}\nonumber\\&&\qquad\times\: \underbrace{\frac{\sin^2[(\mathfrak{w}_n  - \om_n)L/2]\sin^2[(\mathfrak{w}_n  + \om_n)L/2]}{(\mathfrak{w}_n  - \om_n)^2 L^2}}_{\sim\,1/4}\;,
\eea
and everything combines to give the simple result 
\be
\label{eq:electricfinal}
N_n(t) \ =\ 1\: -\: e^{- 2 |\Im(\mathfrak{w}_n)| t}\;,
\ee
which holds only for $t\geq 0$. We see that the modes with $0 < \om_n < V_0$ are initially unoccupied. On lowering the barrier, the vacuum state decays to the true ground state, and these modes become occupied. In the next section, we will return to the Kerr problem, where we will find an analogous result.


\section{Vacuum decay in Kerr}
\label{sec:Kerrvac}

We now return to the original Kerr problem, set up in Sec.~\ref{Kerr_Classical_Sec}. For the electric analogue of Sec.~\ref{sec:electrictoy}, the boundary conditions were changed from those of the initial to the final states by instantaneously lowering the infinite barrier at $t=0$. For the Kerr case, the emergence of the effective potential in Fig.~\ref{potential_Fig} will depend on the way in which the Kerr black hole is formed. One might envisage, e.g., realistic scenarios where the initial condition corresponds to some asymmetrically collapsing mass distribution that forms the Kerr black hole or, alternatively, one might consider an initial Schwarzschild black hole that is spun up by accreting matter from an orbiting body. For our purposes, however, we can be far more pragmatic, since we do not expect the state to depend strongly on the initial conditions at sufficiently late times. What we do expect is that states associated with the upper frequency continuum (see Fig.~\ref{potential_Fig}) are initially empty. This initial state would resemble the `Unruh vacuum' of massless fermions~\cite{Casals12,Ambrus15}. Note that this state normally radiates a thermal flux at the Hawking temperature $T_H$, but because we mainly consider modes $\om \lesssim \mu$, and usually $T_H \ll \mu$, we shall neglect the Hawking temperature. It would, however, be interesting to investigate this in more detail and, in particular, whether thermal effects could somehow accelerate the decay via stimulated emission. 

In order to construct the initial state, we would like to build normalizable modes associated with the relevant QNM frequencies. This is a delicate and ambiguous procedure, since it is a well-known issue of QNMs that they are spatially growing and hence not normalizable~\cite{Zeldovich61,Kokotas99,Coutant16}. Here, we circumvent the problem by constructing approximate eigenstates associated with the real part $\om_n^R$, and of finite support, in close analogy to the preceding section.\footnote{Our construction is very similar to that of `quasi-modes,' which have been considered in the literature~\cite{Zworski17}, and in particular in the context of black holes~\cite{Keir14}.}
 An alternative way of dealing with these modes and constructing a pseudo-orthonormal basis would be to define them as approximations to the exact eigenstates by means of complex distributions~\cite{Nakanishi1,Nakanishi2}. Importantly, we assume that the \emph{in} modes constructed this way form a complete basis, at least in the relevant range of frequencies $\om_\m < \om < \min(\mu, m\Om_H)$, exactly as in equation \eqref{InOut_Field_Decomp} for the electric case. 

To construct the \emph{in} basis explicitly, we first notice that the modes described in Sec.~\ref{QNM_Sec} are a set of trapped modes in the region $r_2<r<r_3$ that can tunnel to the continuum set living in $r<r_1$ (see Fig.~\ref{potential_Fig}). By analogy with the electric case, the tunneling amplitude plays a role similar to the barrier strength $\xi$. There, we can construct approximate \emph{in} modes $\varphi_{n}^{\rm in}$ by assuming that they have support only in the region $r_2<r<r_3$. We select the discrete set satisfying the Bohr-Sommerfeld quantization condition \eqref{QNM_Re} and normalize them using the integral of the scalar product \eqref{eq:sep} between the turning points $r_2$ and $r_3$. In this way, the radial part of the initial and final states are given in the region $r_2<r<r_3$ by 
\be \label{Kerr_modes_in}
R^{\rm in}_n \simeq \frac{1}{\sqrt{2T_{\rm cl}}} \left[ \bmat \frac{i \lam - \mu r}{\sqrt{2 |k_n| (K_{\om_n^R} - \Delta k_n)}} \\ -\sqrt{\frac{K_{\om_n^R} - \Delta k_n}{2 |k_n| \Delta}} \emat e^{+i \int_{r_3}^r \D r'k_n(r') - i\frac{\pi}{4} }\: +\: \bmat \frac{i \lam - \mu r}{\sqrt{2 |k_n| (K_{\om_n^R} + \Delta k_n)}} \\ -\sqrt{\frac{K_{\om_n^R} + \Delta k_n}{2 |k_n| \Delta}} \emat e^{-i \int_{r_3}^r\D r' k_n(r')  + i\frac{\pi}{4}}\right]\;, 
\ee
and 
\be \label{Kerr_modes_out}
R^{\rm out}_\om \ \simeq\ A_\om \left[\bmat \frac{i \lam - \mu r}{\sqrt{4\pi |k| (K_\om - \Delta k)}} \\ -\sqrt{\frac{K_\om - \Delta k}{4\pi |k| \Delta}} \emat e^{+i \int_{r_3}^r  \D r' k(r') - i \frac\pi4}\: +\: \bmat \frac{i \lam - \mu r}{\sqrt{4\pi |k| (K_\om + \Delta k)}} \\ -\sqrt{\frac{K_\om + \Delta k}{4\pi |k| \Delta}} \emat e^{-i \int_{r_3}^r \D r' k(r')  + i \frac\pi4}\right]\;. 
\ee
In the above definition of \emph{in} modes (and of $k_n$), the frequencies are the \emph{real parts} $\om_n^R$ of the QNM frequencies given in equation \eqref{eq:QNMdef}. $T_{\rm cl}$ is defined in equation~\eqref{eq:Tcl}. Notice that the $R^{\rm in}_n$ have been normalized as a discrete set of modes, hence there is a factor $\sqrt{2\pi}$ that differs from equations \eqref{WKB_Ampl} evaluated at $\om \sim \om_n^R$. 

We now show how the initial vacuum, where the states $|\varphi^{\rm in}_{n} \rangle$ are unoccupied, decays spontaneously by filling them in. Proceeding in exact analogy to the electric toy model in Sec.~\ref{sec:electrictoy}, we can define the instantaneous occupation number of the state $|\varphi^{\rm in}_{n} \rangle$ as
\be \label{Kerr_Pnumber}
N_n(t) \ =\ \langle 0^{\rm in}|\hat a_n^\dagger(t) \hat a_n(t) |0^{\rm in} \rangle\ \simeq\ 1\:-\:\left| \int{\rm d}\om\; |\langle \varphi_{n}^{\rm in} | \varphi_{\om}^{\rm out} \rangle|^2 e^{-i \om t} \right|^2\;.
\ee
As before, this occupation number is governed by the analytic structure of the overlap $\braket{\varphi_{n}^{\rm in}|\varphi_{\omega}^{\rm out}}$. Since we work in the WKB regime, we shall estimate this overlap by exploiting the fact that the relevant exponentials oscillate rapidly. By means of the separation in Eq.~\eqref{eq:sep}, we have  
\begin{subequations}
\bea
\braket{\varphi_{n}^{\rm in}|\varphi_{\omega}^{\rm out}}\ & =&\
\int_{r_2}^{r_3}{\rm d}r\;\frac{r^2+a^2}{\Delta}\;R^{\rm in*}_nR^{\rm out}_\om \\
&\simeq&\ \frac{A_{\omega}}{\sqrt{2T_{\rm cl}}}\int_{r_2}^{r_3}{\rm d}r\;\frac{r^2+a^2}{\Delta}\,\big[R_{n,1}^{0*}R_{\omega,1}^0+R_{n,2}^{0*}R_{\omega,2}^0\big]\nonumber\\ &&\qquad \times\:\Big[e^{+i\int{\rm d}r'[k(r')-k_n(r')]}+e^{-i\int{\rm d}r'[k(r')-k_n(r')]}\Big]\;,
\eea
\end{subequations}
where we have neglected the highly-oscillatory and therefore subdominant contributions proportional to $e^{+i\int{\rm d}r'[k(r')+k_n(r')]}$ and $e^{-i\int{\rm d}r'[k(r')+k_n(r')]}$. The remaining integrals are also highly oscillatory except when $\om - \om_n^R \sim 0$, such that the overlap has dominant support for $\omega\sim \om_n^R$. We can therefore expand the phase to leading order in $\om - \om_n^R$. Doing so, we have
\be
\int_{r_3}^r {\rm d}r'\,[k(r')-k_n(r')]\ \simeq\ (\omega-\om_n^R)t_{\rm cl}\;,
\ee
where
\be
t_{\rm cl}\  =\ \int_{r_3}^r\frac{{\rm d}r'}{v_g(r')}
\ee
is the time lapse (in the Boyer-Lindquist coordinate system) necessary to propagate from a reference point (here $r_2$) to $r$. In this way, we obtain
\bea
\braket{\varphi_{n}^{\rm in}|\varphi_{\omega}^{\rm out}}\ & =&\ 2\,\frac{A_{\omega}}{\sqrt{2T_{\rm cl}}}\int_{t_2}^{t_3}{\rm d}t_{\rm cl}\;\frac{r^2+a^2}{\Delta}\,v_g(r)\big[R_{n,1}^{0*}R_{\omega,1}^0+R_{n,2}^{0*}R_{\omega,2}^0\big]\nonumber\\ &&\qquad \times\cos[(\omega-\om_n^R)t_{\rm cl}]\;.
\eea
By estimating the amplitudes $R_{1,2}^0$ at leading order in $\om - \om_n^R$ (which is justified in the WKB approximation), we recognize that $\frac{r^2+a^2}{\Delta} \left[|R_{n,1}^{0}|^2 + |R_{n,2}^{0}|^2\right]$ is nothing else than $1/v_g(r)$. This is a direct consequence of the normalization of the \emph{in} modes $\varphi_n^{\rm in}$, and the correspondence between the norm and the conserved current (see appendix~\ref{Norm_App} and specifically equation \eqref{Current_Norm_density_relation}). This leads to 
\begin{subequations}
\bea
\braket{\varphi_{n}^{\rm in}|\varphi_{\omega}^{\rm out}}\ & \simeq &\ 2\,\frac{A_{\omega}}{\sqrt{\pi T_{\rm cl}}}\,\frac{\sin[(\omega-\om_n^R)T_{\rm cl}/2]\cos[(\omega-\om_n^R)(t_3+t_2)/2]}{\omega-\om_n^R}\\ &\underset{|\om\, -\, \om_n^R| \to 0}{\to}&\ A_{\omega}\sqrt{\frac{T_{\rm cl}}\pi}\;.
\label{eq:kerroverlap}
\eea
\end{subequations}

Now, to evaluate the integral giving the evolution of the occupation number, i.e.~equation \eqref{Kerr_Pnumber}, we need to deform the contour in the lower-half complex $\om$-plane ($t>0$). Doing so, we pick up all the poles of $|A_\om|^2$, in perfect analogy with the electric case. These poles are nothing else than the QNM frequencies, and, as in Sec.~\ref{sec:electrictoy}, only the QNM corresponding to the same index $n$ contributes significantly. For the Kerr case, the residue of $|A_\om|^2$ at its pole is given by
\be
\textrm{Res}(|A_\om|^2; \om_n^R) \ =\ \frac{i}{2 T_{\rm cl}}\;,
\ee
and hence, after making use of Eq.~\eqref{eq:kerroverlap}, we find
\begin{subequations}  
\bea 
\int {\rm d}\omega \; |\langle \varphi_{n}^{\rm in} | \varphi_\om^{\rm out}\rangle|^2 e^{-i \om t}\ &\sim&\ -\,2i\pi\: \times\: \:\times\: \frac{i}{2 T_{\rm cl}}\: \times\: \frac{T_{\rm cl}}\pi \:\times\: e^{-i \mathfrak{w}_n t} \\
&\sim&\ e^{-i \mathfrak{w}_n t}\;. 
\eea
\end{subequations}
By means of equation~\eqref{Kerr_Pnumber}, we then arrive at the simple result, analogous to equation~\eqref{eq:electricfinal}, that the occupation number at the time $t$ is given by
\be
N_n(t) \ =\ 1\: -\: e^{- 2 |\Im(\mathfrak{w}_n)| t}\;. 
\ee

This shows what was announced earlier: the initial vacuum slowly decays by filling in the discrete set of energy levels corresponding to the QNM satisfying $\om^R < \min(\mu, m\Om_H)$. Since these modes are trapped, this process is not accompanied by any radiation emitted to infinity. The decay process we describe does not arise on top of the Unruh-Starobinski radiation, but it replaces it. In fact, we can decompose the process in three steps. First, following initial transients, the field is in the non-rotating vacuum. Then, as long as the exponentials $e^{-i \mathfrak{w}_n t}$ are approximately linear in $t$, there is a steady flux of radiation emitted from the ergoregion: the Unruh-Starobinski radiation. This radiation fills in the discrete set of trapped modes, until every level is filled and the process stops. One is left with a Fermi sea, as described in Ref.~\cite{Hartman09}, which now possesses a non-zero expectation value of the angular momentum. To see this, we can evaluate the total angular momentum inside the trapped region. Assuming that the \emph{in} modes used earlier are complete (at least approximately), we have 
\bsub \bea
\langle \hat L_{\rm trapped}(t) \rangle \ &\doteq&\ \langle \int_{r_2}^{r_3} \D r \oint \D \Om \;\hat T_{t \phi} \rangle \approx \sum_{0\, <\, \om_n^R\, <\, \m(\mu, m\Om_H)} m \langle (\hat a^{\rm in}_n)^\dagger \hat a^{\rm in}_n \rangle \\
&\approx&\ \sum_{0\, <\, \om_n^R\, <\, \m(\mu, m\Om_H)} m \left(1\: -\: e^{- 2 |\Im(\mathfrak{w}_n)| t}\right)\;. \label{L_decay}
\eea \esub 
Notice that the above sum runs over the (discrete) set of QNMs satisfying the condition $\om^R\, <\, \m(\mu, m\Om_H)$. In particular, we sum over the quantum numbers $j$ and $\mathcal P$, thereby taking into account the contributions of all values of the angular momentum and the spin projections. Note also that the stress-energy tensor is renormalized at early times, which amounts to normal ordering with respect to the $\hat a^{\rm in}$'s. We then conclude that the true vacuum possesses a mean angular momentum given by 
\be \label{L_vev}
\langle \hat L_{\rm trapped}(t \to \infty) \rangle\ =\ \sum_{0\, <\, \om_n^R\, <\, \m(\mu, m\Om_H)} m\;. 
\ee
Of course, if one computes the total angular momentum on the whole space, one finds that it is conserved and vanishes. This means that the spontaneous acquisition of an angular momentum vacuum expectation value is accompanied by a net flux in the horizon of the black hole, thereby slowing it down, as in bosonic superradiant processes. Although the total angular momentum \eqref{L_vev} can be a significant fraction of the black hole angular momentum (see section 3.1.2 in Ref.~\cite{Hartman09}) depending on the parameter $M\mu$, after a finite time, only a fraction of states have decayed. The extraction of angular momentum from the black hole is mainly a collective effect, since each state carries a small angular momentum. Hence, a proper estimate of the angular momentum transfer after a time $t$ requires us to work with equation \eqref{L_decay} rather than \eqref{L_vev}.

To finish this section, we point out that we have carried out this analysis in the regime $M \mu \gg 1$, wherein the time necessary for the decay is exponentially long (see equation~\eqref{QNM_Im}). However, the same effect exists for any value of the parameter $M\mu$. The modes that contribute to the effect are those satisfying $|\om^I| \ll |\om^R|$ (which justifies the construction of the initial vacuum) and the condition $\om^R < \min(\mu, m\Om_H)$. In the opposite regime, when $M\mu \ll 1$, the lifetime is also large, increasing as a power law in $M\mu$ (see the works of Refs.~\cite{Ternov78,Ternov88}). Based on the results of Ref.~\cite{Dolan09}, the effect is expected to be the quickest in the regime $M\mu \sim 1$, in close similarity to the bosonic black hole bomb instability.

\section{Conclusion}

In this work, we have shown that massive fermions around a rotating black hole are subject to a vacuum instability. After the black hole is formed, the initial state is similar to the Unruh vacuum and has a zero angular momentum expectation value. As time progresses, this state decays into the true vacuum, which corresponds to the Kerr-Fermi sea identified by Hartman {\it et al.}~\cite{Hartman09}. We show that this decay is entirely governed by the set of QNMs satisfying the superradiance condition $0 < \om < m\Om_H$ (although there is no amplification) and lying below the mass gap $\om < \mu$. These modes correspond to a discrete set of bound states, which orbit well outside the ergoregion. The states are progressively filled with particles, with an exponential decay probability characterized by the imaginary part of the QNM frequency. Once every state is filled with one particle of each spin, the process stops. As usual, the vacuum decay is accompanied by a spontaneous symmetry breaking. Here, the true vacuum acquires a non-zero expectation value of angular momentum. 

When the effect is applied to particles of the Standard Model, the time scale of the decay is extremely large. Using a conservative estimate for the mass of the light neutrino, and a solar mass black hole, one has $M \mu \sim 10^9$, and therefore, the imaginary parts of the QNM frequencies are suppressed by a factor $e^{-10^9}$, making the decay time larger than the age of the Universe. This suggests that the state of neutrino fields around astrophysical black holes is likely to be the metastable, non-rotating one. However, this statement deserves a more precise analysis. Indeed,  since the density of states is also very large in the limit $M \mu \gg 1$, by statistical effects, several states will be filled after a reasonable amount of time. It would be interesting to study the structure of the state as a function of time, when only a fraction of the bound states has been filled. Moreover, since the decay is accompanied by a decrease of the black hole angular momentum, this effect could be used to constrain the possible range of mass for light fermions, in the same spirit as was done for light bosons using the black hole bomb instability~\cite{Arvanitaki09,Pani12,Brito13,Brito17}. One can also imagine various alternative scenarios where the quantity $M \mu$ becomes of order 1, and the vacuum instability becomes relevant. This could happen around a primordial black hole with a sufficiently small mass or in the presence of ultra-light fermions~\cite{Arvanitaki09}. In this context, it is interesting to remark that the mass of the lightest SM neutrino does not have a lower experimental or observational bound~\cite{Tanabashi18}.

In this work, we have concentrated on Kerr black holes. A similar effect is anticipated to be present also for charged black holes, as described by the Reissner-Nordstr\"{o}m spacetime. In this case, it is expected that the true vacuum will acquire a non-zero charge instead of angular momentum, as in the electric model of Sec.~\ref{sec:electrictoy}. This vacuum instability could also be of interest for holography, where similar effects have been studied in the context of holographic superconductivity~\cite{Lee08,Liu09}, or for the Kerr-CFT correspondence~\cite{Guica08}.

\begin{acknowledgments}

We thank Helvi Witek for early discussions on the problem, and Vitor Cardoso, Sam Dolan, Renaud Parentani, Harvey Reall and Jorge Santos for interesting and helpful comments. AC has received fundings from the European Union's Horizon 2020 Research and Innovation Programme under the Marie Sk\l odowska-Curie Grant Agreement No.~655524, and the Nottingham Advanced Research Fellowship (A2RHS2). The work of PM was supported by the Science and Technology Facilities Council (STFC) under Grant No.~ST/L000393/1 and a Leverhulme Trust Research Leadership Award. 

\end{acknowledgments}

\newpage
\appendices

\section{Dirac equation on a curved background in the Weyl representation}
\label{Dirac_CST_App}

For completeness, we provide in this appendix all the conventions for the Dirac equation in Kerr, as used in this work. We recall that we have chosen the same conventions as in Ref.~\cite{Dolan15} and we refer to the literature on the subject for more details~\cite{Chandrasekhar}. To define the Dirac equation \eqref{D_eq} fully, we need to build the spinor covariant derivative $D_\mu$ and the curved-spacetime Dirac matrices $\gamma^\mu$. To do so, we use the tetrad formalism. We first write the metric as $g_{\mu \nu} e^{\mu}_a e^{\nu}_b = \eta_{ab}$, where $\eta_{ab}$ is the Minkowski metric (see equation (11) in Ref.~\cite{Dolan15} for the explicit expressions in Kerr). We use lower-case Greek characters to label coordinates in the curved space and lower-case Roman characters to label those in the tangent space.

Using the tetrad $e_a^{\mu}$, we construct the curved-spacetime Dirac matrices $\gamma^\mu$ from their flat-spacetime counterparts $\hat \gamma^\mu$: 
\be
\gamma^\mu\ =\ e^\mu_a \hat \gamma^a\;. 
\ee
Notice that $\gamma^\mu$ and $e^\mu_a$ depend on the point in the spacetime manifold, whereas the $\hat \gamma^a$ do not. In the Weyl/Chiral representation, the flat Dirac matrices are given by 
\be
\hat{\gamma}^0 \ =\ \bmat 0 & I_2 \\ I_2 & 0 \emat \qquad \textrm{and} \qquad \hat{\gamma}^j \ =\ \bmat 0 & \sigma^{j} \\ -\sigma^{j} & 0 \emat\;, \qquad j=1,2,3\;, 
\ee
where $\sigma^j$ ($j=1,2,3$) are the Pauli matrices 
\be
\sigma^{1} \ =\ \bmat 0 & 1 \\ 1 & 0 \emat\;, \quad \sigma^{2}\ =\ \bmat 0 & -i \\ i & 0 \emat\;, \quad \sigma^{3} \ =\  \bmat 1 & 0 \\ 0 & -1 \emat\;.
\ee
The spinor covariant derivative can be written  
\be 
D_\nu\ =\ \p_\nu \: +\: \frac14 \om_{\nu\; ab} \hat \gamma^a \hat \gamma^b\;, 
\ee
where 
\be
\om_{\nu \; ab}\ =\ \frac12 e_{\nu}^c \left(\lam_{abc} + \lam_{cab} - \lam_{bca}\right)
\ee
is the spin connection in which $\lam_{abc} = e^\mu_a(\p_\nu e_{b\mu} - \p_\mu e_{b\nu}) e^\nu_c$.


\section{Eigenmode normalization and conserved current}
\label{Norm_App}
In this appendix, we show that the normalization of the conserved norm \eqref{modes_norm} is equivalent to that of the current \eqref{current_norm}. We shall present a very general argument.

Let's consider a field $\phi: \mathbb R^2 \to \mathbb C^n$ in 1+1 dimensions $(t,x)$. Associated with the dynamics of the field is a conservation law of the form 
\be \label{Conservation_law}
\p_t D[\phi]\: +\: \p_x J[\phi]\ =\ 0\;, 
\ee
where the norm density $D$ and the current $J$ are quadratic forms. In the core of the paper, $\mathbb C^n$ is the space of the four components of a Dirac spinor, and the radial coordinate $r$ plays the role of $x$, once one has integrated over the angles. The conservation law above corresponds to equation \eqref{Kerr_conservation}. One can already draw two important consequences of equation \eqref{Conservation_law}: 
\ben
\item If $\phi(t,x)$ is localized in space, that is $\phi(t,.) \in \mathbb C^n \otimes L^2(R_x)$, then integration of equation \eqref{Conservation_law} over space implies that the norm is conserved in time, i.e. 
\be \label{App_Norm_Cons}
\int {\rm d}x\; D[\phi]\ =\ \text{const\;.} 
\ee
\item If $\phi(t,x)$ is a stationary mode, meaning $\phi(t,x) = \varphi_\om(x) e^{-i \om t}$ for some real frequency $\om$, then the current $J[\varphi_\om(x)]$ is conserved along $x$. 
\een

Now, we assume that any solution of the equation of motion that is localized in space can be written as a \emph{dense} sum of eigenfrequency modes $\varphi_\om$. For simplicity, we assume that the background is homogeneous, and hence the eigenmodes are plane waves: 
\be \label{App_Plane_waves}
\varphi_\om(x) \ =\ C_\om e^{i k_\om x}\;,  
\ee
where $C_\om$ is a vector in the internal space $\mathbb C^n$. However, the same result applies directly to WKB modes, simply by computing the various integrals using appropriate stationary phase approximations. We want to normalize the eigenbasis in the Dirac sense, that is\footnote{Of course, in this equation $D[\varphi_\om, \varphi_{\om'}]$ is evaluated using the sesquilinear form canonically associated with the quadratic form $D$. Also, the fact that modes of different frequency are orthogonal with respect to $D$ (which justifies the $\delta$-function in equation~\eqref{App_Norm}) comes from the norm conservation in time \eqref{App_Norm_Cons} and the reality of the frequencies (see, e.g., Refs.~\cite{Fulling,Coutant16} in the context of curved spacetimes).}
\be \label{App_Norm}
\int {\rm d}x\; D[\varphi_\om, \varphi_{\om'}]\ =\ N_\om \delta(\om - \om')\;, 
\ee
with $N_\om = \pm 1$. The normalization conditions generally fix the amplitude $C_\om$ in equation \eqref{App_Plane_waves}. Our general claim is that this normalization is equivalent to that of the conserved current. More specifically, we will show that 
\be \label{Norm_Theorem}
|N_\om| \ =\ 1\ \Leftrightarrow\ |J[C_\om]|\ =\ \frac1{2\pi}\;, 
\ee
which is what was used in the main text. For this, we fix a frequency $\om$ and consider a tight wave packet whose Fourier spectrum is peaked around $\om$, i.e. 
\be
\phi(t,x)\ =\ \int {\rm d}\om'\; f(\om' - \om) \varphi_{\om'}(x) e^{-i \om' t}\;, 
\ee
where the function $f$ is highly peaked about 0. 

The usual trick to describe tight wave packets is to consider only the vicinity of $\om$. For this, we define $\om' = \om + \eps$ and expand phases to first order in $\eps$:
\bsub \bea
\phi(t,x)\ &\sim&\ C_\om e^{-i \om t + i k_\om x} \int {\rm d}\eps\; f(\eps) e^{-i \eps t + i \eps x/v_g} \\
 &\sim&\ C_\om e^{-i \om t + i k_\om x} \hat f\left(t - x/v_g\right)\;, 
\eea \esub
where $\hat f$ is the (inverse) Fourier transform of $f$. We recover here the standard result: $\phi$ essentially oscillates as $e^{-i \om t + i k_\om x}$, and since $f$ is sharply peaked, $\hat f$ is a long envelope (i.e.~containing many wavelengths defined by $k_\om$), moving at the group velocity $v_g = (\p_\om k)^{-1}$. Notice that, if one works with WKB modes the argument of $f$ is replaced by $t - \int {\rm d}x/v_g(x)$ since the group velocity is defined locally. In the case of a field obeying a wave equation in a curved spacetime, this exactly means that tight wave packets follow null geodesics. 

We will now use this expression in the conservation law \eqref{Conservation_law}. Noticing that   
\be
D\left[\phi(t,x)\right] \ =\ \left|\hat f\left(t - x/v_g\right)\right|^2 D[C_\om] \qquad \textrm{and} \qquad J\left[\phi(t,x)\right] \ = \ \left|\hat f\left(t - x/v_g\right)\right|^2 J[C_\om]\;, 
\ee
it follows that
\be
2 \Re\left(\hat f^* \hat{f}'\right) D[C_\om] \:-\: \frac2{v_g}\, \Re\left(\hat f^* \hat{f}'\right) J[C_\om] \ =\ 0\;. 
\ee
Hence, we obtain the following identity, valid for any $\om$: 
\be \label{Current_Norm_density_relation}
J[C_\om]\ =\ v_g D[C_\om]\;. 
\ee
This is the natural translation of the conservation law for a definite frequency: the current is simply the norm density times the moving velocity of wave packets $v_g$. All that is left to do now is to use it in the normalization condition \eqref{App_Norm}: 
\bsub \bea
N_\om \delta(\om - \om')\ &=&\ \int {\rm d}x\; D[\varphi_\om, \varphi_{\om'}] \\
&=&\ D[C_\om, C_{\om'}] \int {\rm d}x\; e^{i (k_{\om'} - k_{\om}) x} \\ 
&=&\ D[C_\om] 2\pi \delta(k_{\om'} - k_{\om}) \\
&=&\ 2\pi D[C_\om] |v_g| \delta(\om - \om')\;. 
\eea \esub
Using the identify \eqref{Current_Norm_density_relation}, we have 
\be
N_\om \ =\  2\pi \sign(v_g) J[C_\om]\;,
\ee
leading directly to the result in equation~\eqref{Norm_Theorem}.

\section{Connection formula for the Dirac equation}
\label{app:turning}

In this appendix, we briefly review the reasoning that leads to the boundary conditions used at the turning points to build the modes in Kerr (see equations \eqref{Kerr_modes_left} and \eqref{Kerr_modes_right}). The idea is rather standard from WKB methods~\cite{Berry72,Gottfried} and amounts to determining how to connect two oscillating solutions on one side of the turning point to the growing and decaying exponentials on the other side. To see this, we consider a general solution of the left of the outer turning point $r_3$. For $r_2 < r < r_3$, it is a sum of oscillating exponentials obtained from equations \eqref{WKB_ansatz} and \eqref{WKB_Ampl}: 
\be
R(r)\ = \
a \bmat \frac{i \lam - \mu r}{\sqrt{4\pi |k| (K_\om - \Delta k)}} \\ -\sqrt{\frac{K_\om - \Delta k}{4\pi |k| \Delta}} \emat e^{+i \int_{r_3}^r \D r'k(r') }\: +\: b \bmat \frac{i \lam - \mu r}{\sqrt{4\pi |k| (K_\om + \Delta k)}} \\ -\sqrt{\frac{K_\om + \Delta k}{4\pi |k| \Delta}} \emat e^{-i \int_{r_3}^r\D r' k(r')}\;. 
\ee
Near the turning point $r_3$, $k\sim \alpha |r-r_3|^{1/2}$. Writing $\varepsilon = r-r_3$, we have 
\be
R(r)\ \sim\ 
\bmat \frac{i \lam - \mu r}{\sqrt{K_\om}} \\ -\sqrt{\frac{K_\om}{\Delta}} \emat \left(a \frac{e^{i \frac{2\alpha}3 |\varepsilon|^{3/2} }}{\sqrt{4\pi\alpha |\varepsilon|^{1/2}}}\: +\: b \frac{e^{-i \frac{2\alpha}3 |\varepsilon|^{3/2} }}{\sqrt{4\pi\alpha |\varepsilon|^{1/2}}} \right)\;. 
\ee
We see that this combination is the same as the usual case of a scalar function. The usual trick is to identify it with the asymptotic behaviour of an Airy function, which furnishes us with a regular solution across the turning point. Using the following asymptotics of the Airy function:
\be
\frac{e^{-i \frac23 (-z)^{\frac32} + i\frac{\pi}4}}{2|z|^{\frac14}\sqrt{\pi}}\: +\: \frac{e^{i \frac23 (-z)^{\frac32} - i\frac{\pi}4}}{2|z|^{\frac14}\sqrt{\pi}}\ \underset{-\infty}\leftarrow\ \Ai(z)\ \underset{+\infty}{\to}\ \frac{e^{- \frac23 z^{\frac32}}}{2|z|^{\frac14}\sqrt{\pi}}\;, 
\ee
we conclude that 
\be
b\ =\ a e^{i \frac\pi2}\;, 
\ee
and 
\be 
R(r)\ \sim\ 
\bmat \frac{i \lam - \mu r}{\sqrt{K_\om}} \\ -\sqrt{\frac{K_\om}{\Delta}} \emat \left(a\,\frac{e^{-\frac{2\alpha}3 |\varepsilon|^{3/2} - i \frac\pi4}}{\sqrt{4\pi\alpha |\varepsilon|^{1/2}}} \right) \qquad \textrm{for}  \qquad r\ >\ r_3\;. 
\ee

In order to obtain the Kerr modes of equations \eqref{Kerr_modes_left} and \eqref{Kerr_modes_right}, one needs to apply the connection formula three times: at $r_1$, $r_2$ and $r_3$. Notice that we have to face the standard problem of the irreversibility of the connection formula~\cite{Berry72}. However, as we saw, everything happens exactly as in the scalar case, exposed in detail in most treatise, and one can circumvent the irreversibility problem in the same way. As a final remark, we point out that what we used in this work is a combination of (three times) the single turning point formula. It is possible to obtain more general three-turning-point connection formulae~\cite{Berry72}. Doing so is significantly more involved and barely improves the accuracy of our results.

\section{Further details of the electric toy model}
\label{Eletric_App}
In this appendix, we include further details for the electric toy model in Sec.~\ref{sec:electrictoy}, including the complete mode functions, their overlaps and the decomposition of the field operator.

We first compute the set of four linearly independent solutions. For $x < -D$, we have $V(x) = V_0$, and $\mu(x) = 0$. This gives four solutions 
\bsub \bea
\psi_1^L\ =\ \bmat 1 \\ 0 \\ 0 \\ 0 \emat e^{-i (\om - V_0)x}\;, &\qquad& \psi_2^L \ =\ \bmat 0 \\ 1 \\ 0 \\ 0 \emat e^{+i (\om - V_0)x}\;, \\
\psi_3^L\ =\ \bmat 0 \\ 0 \\ 1 \\ 0 \emat e^{+i (\om - V_0)x}\;, &\qquad& \psi_4^L \ =\ \bmat 0 \\ 0 \\ 0 \\ 1 \emat e^{-i (\om - V_0)x}\;. 
\eea \esub
Similarly, for $D < x < L$, we have $V(x)=0$ and 
\bsub \bea
\psi_1^R\ =\ \bmat 1 \\ 0 \\ 0 \\ 0 \emat e^{-i \om x}, &\qquad& \psi_2^R \ =\ \bmat 0 \\ 1 \\ 0 \\ 0 \emat e^{+i \om x}\;, \\
\psi_3^R\ =\ \bmat 0 \\ 0 \\ 1 \\ 0 \emat e^{+i \om x}, &\qquad& \psi_4^R \ = \ \bmat 0 \\ 0 \\ 0 \\ 1 \emat e^{-i \om x}\;. 
\eea \esub 
Note that the subscripts $L$ and $R$, indicate the solutions to the left and right of the barrier and not chiralities. Lastly, inside the barrier, we have $V(x) = 0$ and $\mu(x) = \mu_0$. We assume that $|\om| < \mu_0$. (We will ultimately take the limit $\mu_0 \to \infty$.) In this case, the modes are either growing or decaying but not oscillating. Defining $\kappa = \sqrt{\mu_0^2 - \om^2} > 0$, we get 
\bsub \bea
\psi_1^I \ =\  \bmat \mu_0 \\ 0 \\ \om - i \kappa \\ 0 \emat e^{- \kappa x}\;, &\qquad& \psi_2^I \ =\ \bmat 0 \\ \mu_0 \\ 0 \\ \om + i \kappa \emat e^{- \kappa x}\;, \\
\psi_3^I \ =\  \bmat \mu_0 \\ 0 \\ \om + i \kappa \\ 0 \emat e^{+\kappa x}\;, &\qquad& \psi_4^I\ =\ \bmat 0 \\ \mu_0 \\ 0 \\ \om - i \kappa \emat e^{+\kappa x}\;. 
\eea \esub
Using these and assuming continuity of the field for the potential and mass of Fig.~\ref{Elec_potential_Fig}, we can extract from them the boundary conditions \eqref{BC_elec1} and \eqref{BC_elec2}. The outgoing modes are given by a single continuum, composed of  
\bea
\varphi_{\om}^{\rm out}(x)\ &=&\ \begin{cases} \frac1{\sqrt{2\pi}} \bmat e^{-i (\om - V_0) x + i \pi/4 + 2 i \delta_\om} \\ 0 \\ e^{+i (\om - V_0) x - i \pi/4} \\ 0 \emat\;, &\qquad x\ <\ 0\;, \\
\frac{A_\om}{\sqrt{2L}} \bmat e^{-i \om (x-L) + i \pi/4} \\ 0 \\ e^{+i \om (x-L) - i \pi/4} \\ 0 \emat\;,& \qquad 0\ <\ x\ <\ L\;.\end{cases} 
\eea 

For completeness, we also give the overlaps of the various mode functions on the interval $[0,L]$: 
\begin{subequations}
\begin{gather}
\label{eq:dominant}
\int_{0}^{L}\!{\rm d}x\;\big(\varphi_{n}^{\rm in}(x)\big)^{\dag}\varphi_{\omega}^{\rm out}(x)\ =\ -\:A_{\omega}\,{\rm sinc}\,\big[(\omega-\omega_{n})\tfrac{L}{2}\big]\sin\big[(\omega+\omega_{n})\tfrac{L}{2}\big]\;,\\
\int_{0}^{L}\!{\rm d}x\;\big(\varphi_{n}^{\rm in}(x)\big)^{\dag}\psi_{\omega}^{\rm out}(x)\ =\ A_{2m\Omega_H-\omega}\,{\rm sinc}\,\big[(\omega+\omega_{n}-2m\Omega_H)\tfrac{L}{2}]\sin\big[(\omega-\omega_{n}-2m\Omega_H)\tfrac{L}{2}\big]\;,\\
\int_{0}^{L}\!{\rm d}x\;\big(\psi_{n}^{\rm in}(x)\big)^{\dag}\varphi_{\omega}^{\rm out}(x)\ =\ -\:A_{\omega}\,{\rm sinc}\,\big[(\omega+\omega_{n})\tfrac{L}{2}\big]\sin\big[(\omega-\omega_{n})\tfrac{L}{2}]\;,\\
\int_{0}^{L}\!{\rm d}x\;\big(\psi_{n}^{\rm in}(x)\big)^{\dag}\psi_{\omega}^{\rm out}(x)\ =\ A_{2m\Omega_H-\omega}\,{\rm sinc}\,\big[(\omega-\omega_{n}-2m\Omega_H)\tfrac{L}{2}\big]\sin\big[(\omega+\omega_{n}-2m\Omega_H)\tfrac{L}{2}\big]\;.
\end{gather}
\end{subequations}
The dominant support of the overlaps occurs at the zeros of the arguments of the ${\rm sinc}$ functions, setting $\omega \sim \omega_n$.

\bibliographystyle{utphys}
\bibliography{Bibli}

\end{document}